\documentstyle[twoside,epsfig]{article}
\newcommand{\Kstar}{\overline{K}^{\;\!*0}}
\newcommand{\cerenkov}{\v{C}erenkov}
\newcommand{\etal}{{\em et al.}}

\def\issue(#1,#2,#3){{\bf #1} (#3) #2} 
\def\PRL(#1,#2,#3){Phys.\ Rev.\ Lett.\ \issue(#1,#2,#3)}
\def\PL(#1,#2,#3){Phys.\ Lett.\ \issue(#1,#2,#3)}
\def\APP(#1,#2,#3){Acta Phys.\ Polon.\ \issue(#1,#2,#3)}
\def\MPL(#1,#2,#3){Mod.\ Phys.\ Lett.\ \issue(#1,#2,#3)}
\def\PRD(#1,#2,#3){Phys.\ Rev.\  \issue(#1,#2,#3)}
\def\CJP(#1,#2,#3){Chin.\ J.\ Phys. (Taipei)\ \issue(#1,#2,#3)}
\def\EPJC(#1,#2,#3){Eur.\ Phys.\ J.\ C\ \issue(#1,#2,#3)}
\def\NIM(#1,#2,#3){Nucl.\ Instrum.\ Meth.\ \issue(#1,#2,#3)}
\def\IEEETNS(#1,#2,#3){IEEE Trans.\ NS \issue(#1,#2,#3)}
\def\EPJD(#1,#2,#3){EPJdirect \issue(#1,#2,#3)}
\def\CPC(#1,#2,#3){Comp.\ Phys.\ Comm.\ \issue(#1,#2,#3)}
\def\ARNPS(#1,#2,#3){Ann.\ Rev.\ Nucl.\ Part.\ Sci.\ \issue(#1,#2,#3)}
\def\CIP(#1,#2,#3){Comput.\ Phys.\ \issue(#1,#2,#3)}
\def\NP(#1,#2,#3){Nucl.\ Phys.\ \issue(#1,#2,#3)}
\def\SJNP(#1,#2,#3){Sov.\ J. Nucl.\ Phys.\ \issue(#1,#2,#3)}
\def\ZPC(#1,#2,#3){Zeit.\ Phys.\ C \issue(#1,#2,#3)}

\catcode`\@=11
\long\def\@makefntext#1{
\protect\noindent \hbox to 3.2pt {\hskip-.9pt  
$^{{\eightrm\@thefnmark}}$\hfil}#1\hfill}		

\def\@makefnmark{\hbox to 0pt{$^{\@thefnmark}$\hss}}	
	
\def\ps@myheadings{\let\@mkboth\@gobbletwo
\def\@oddhead{\hbox{}
\rightmark\hfil\eightrm\thepage}   
\def\@oddfoot{}\def\@evenhead{\eightrm\thepage\hfil
\leftmark\hbox{}}\def\@evenfoot{}
\def\sectionmark##1{}\def\subsectionmark##1{}}

\oddsidemargin=\evensidemargin
\newcounter{sectionc}\newcounter{subsectionc}\newcounter{subsubsectionc}
\renewcommand{\section}[1] {\vspace{12pt}\addtocounter{sectionc}{1} 
\setcounter{subsectionc}{0}\setcounter{subsubsectionc}{0}\noindent 
	{\tenbf\thesectionc. #1}\par\vspace{5pt}}
\renewcommand{\subsection}[1] {\vspace{12pt}\addtocounter{subsectionc}{1} 
	\setcounter{subsubsectionc}{0}\noindent 
	{\bf\thesectionc.\thesubsectionc. {\kern1pt \bfit #1}}\par\vspace{5pt}}
\renewcommand{\subsubsection}[1] {\vspace{12pt}\addtocounter{subsubsectionc}{1}
	\noindent{\tenrm\thesectionc.\thesubsectionc.\thesubsubsectionc.
	{\kern1pt \tenit #1}}\par\vspace{5pt}}
\newcommand{\nonumsection}[1] {\vspace{12pt}\noindent{\tenbf #1}
	\par\vspace{5pt}}

\newcounter{appendixc}
\newcounter{subappendixc}[appendixc]
\newcounter{subsubappendixc}[subappendixc]
\renewcommand{\thesubappendixc}{\Alph{appendixc}.\arabic{subappendixc}}
\renewcommand{\thesubsubappendixc}
	{\Alph{appendixc}.\arabic{subappendixc}.\arabic{subsubappendixc}}

\renewcommand{\appendix}[1] {\vspace{12pt}
        \refstepcounter{appendixc}
        \setcounter{figure}{0}
        \setcounter{table}{0}
        \setcounter{lemma}{0}
        \setcounter{theorem}{0}
        \setcounter{corollary}{0}
        \setcounter{definition}{0}
        \setcounter{equation}{0}
        \renewcommand{\thefigure}{\Alph{appendixc}.\arabic{figure}}
        \renewcommand{\thetable}{\Alph{appendixc}.\arabic{table}}
        \renewcommand{\theappendixc}{\Alph{appendixc}}
        \renewcommand{\thelemma}{\Alph{appendixc}.\arabic{lemma}}
        \renewcommand{\thetheorem}{\Alph{appendixc}.\arabic{theorem}}
        \renewcommand{\thedefinition}{\Alph{appendixc}.\arabic{definition}}
        \renewcommand{\thecorollary}{\Alph{appendixc}.\arabic{corollary}}
        \renewcommand{\theequation}{\Alph{appendixc}.\arabic{equation}}
        \noindent{\tenbf Appendix \theappendixc #1}\par\vspace{5pt}}
\newcommand{\subappendix}[1] {\vspace{12pt}
        \refstepcounter{subappendixc}
        \noindent{\bf Appendix \thesubappendixc. {\kern1pt \bfit #1}}
	\par\vspace{5pt}}
\newcommand{\subsubappendix}[1] {\vspace{12pt}
        \refstepcounter{subsubappendixc}
        \noindent{\rm Appendix \thesubsubappendixc. {\kern1pt \tenit #1}}
	\par\vspace{5pt}}

\newcommand{\textlineskip}{\baselineskip=13pt}
\newcommand{\smalllineskip}{\baselineskip=10pt}

\def\eightcirc{
\begin{picture}(0,0)
\put(4.4,1.8){\circle{6.5}}
\end{picture}}
\def\eightcopyright{\eightcirc\kern2.7pt\hbox{\eightrm c}}

\def\abstracts#1#2#3{{
	\centering{\begin{minipage}{4.5in}\baselineskip=10pt\footnotesize
	\parindent=0pt #1\par 
	\parindent=15pt #2\par
	\parindent=15pt #3
	\end{minipage}}\par}} 

\renewenvironment{thebibliography}[1]
	{\frenchspacing
	 \ninerm\baselineskip=11pt
	 \begin{list}{\arabic{enumi}.}
	{\usecounter{enumi}\setlength{\parsep}{0pt}
	 \setlength{\leftmargin 12.7pt}{\rightmargin 0pt} 
	 \setlength{\itemsep}{0pt} \settowidth
	{\labelwidth}{#1.}\sloppy}}{\end{list}}

\newcounter{itemlistc}
\newcounter{romanlistc}
\newcounter{alphlistc}
\newcounter{arabiclistc}

\newcommand{\fcaption}[1]{
        \refstepcounter{figure}
        \setbox\@tempboxa = \hbox{\footnotesize Fig.~\thefigure. #1}
        \ifdim \wd\@tempboxa > 5in
           {\begin{center}
        \parbox{5in}{\footnotesize\smalllineskip Fig.~\thefigure. #1}
            \end{center}}
        \else
             {\begin{center}
             {\footnotesize Fig.~\thefigure. #1}
              \end{center}}
        \fi}

\newcommand{\tcaption}[1]{
        \refstepcounter{table}
        \setbox\@tempboxa = \hbox{\footnotesize Table~\thetable. #1}
        \ifdim \wd\@tempboxa > 5in
           {\begin{center}
        \parbox{5in}{\footnotesize\smalllineskip Table~\thetable. #1}
            \end{center}}
        \else
             {\begin{center}
             {\footnotesize Table~\thetable. #1}
              \end{center}}
        \fi}

\def\@citex[#1]#2{\if@filesw\immediate\write\@auxout
	{\string\citation{#2}}\fi
\def\@citea{}\@cite{\@for\@citeb:=#2\do
	{\@citea\def\@citea{,}\@ifundefined
	{b@\@citeb}{{\bf ?}\@warning
	{Citation `\@citeb' on page \thepage \space undefined}}
	{\csname b@\@citeb\endcsname}}}{#1}}

\newif\if@cghi
\def\cite{\@cghitrue\@ifnextchar [{\@tempswatrue
	\@citex}{\@tempswafalse\@citex[]}}
\def\citelow{\@cghifalse\@ifnextchar [{\@tempswatrue
	\@citex}{\@tempswafalse\@citex[]}}
\def\@cite#1#2{{$\null^{#1}$\if@tempswa\typeout
	{IJCGA warning: optional citation argument 
	ignored: `#2'} \fi}}

\def\pmb#1{\setbox0=\hbox{#1}
	\kern-.025em\copy0\kern-\wd0
	\kern.05em\copy0\kern-\wd0
	\kern-.025em\raise.0433em\box0}

\def\fnt#1#2{\footnotetext{\kern-.3em
	{$^{\mbox{\scriptsize #1}}$}{#2}}}

\def\fpage#1{\begingroup
\voffset=.3in
\thispagestyle{empty}\begin{table}[b]\centerline{\footnotesize #1}
	\end{table}\endgroup}

\def\runninghead#1#2{\pagestyle{myheadings}
\markboth{{\protect\footnotesize\it{\quad #1}}\hfill}
{\hfill{\protect\footnotesize\it{#2\quad}}}}
\font\tenrm=cmr10
\font\tenit=cmti10 
\font\tenbf=cmbx10
\font\bfit=cmbxti10 at 10pt
\font\ninerm=cmr9

\font\eightrm=cmr8

\def\qed{\hbox{${\vcenter{\vbox{			
   \hrule height 0.4pt\hbox{\vrule width 0.4pt height 6pt
   \kern5pt\vrule width 0.4pt}\hrule height 0.4pt}}}$}}

\begin{document}

\runninghead{Search for Rare 3 and 4-Body D$^{\,0}$ Decays
$\ldots$} {Search for Rare 3 and 4-Body D$^{\,0}$
Decays $\ldots$}

\normalsize\textlineskip
\thispagestyle{empty}
\setcounter{page}{1}

\begin{table}[t!]
\vspace*{-2.5cm}
\smalllineskip
\tabcolsep=0.0pt
\begin{tabular}{lcl}
\footnotesize International Journal of Modern Physics A 
& \hspace*{20mm} & \footnotesize  UMS/HEP/2000-031 \\
\footnotesize $\eightcopyright$\, World Scientific Publishing Company 
& & \footnotesize  FERMILAB-Conf-00/292-E \\
\footnotesize Division of Particles and Fields & 
& \footnotesize  DPF2000 \ \ \ 9-12 August 2000 \\
\footnotesize American Physical Society 
& & \footnotesize The Ohio State University--Columbus  \\ 
\end{tabular}
\end{table}


\vspace*{0.10truein}

\fpage{1}
\centerline{\bf SEARCH FOR RARE 3 AND 4--BODY D$^{\,0}$ DECAYS AT FNAL E791}
\vspace*{0.37truein}
\centerline{\footnotesize D.~J.~SUMMERS\footnote{Representing the Fermilab E791
Collaboration and supported by DE-FG05-91ER40622.}}
\vspace*{0.015truein}
\centerline{\footnotesize\it Department of Physics and Astronomy, 
University of Mississippi-Oxford} 
\baselineskip=10pt
\centerline{\footnotesize\it University, Mississippi 38677, USA} 
\vspace*{0.21truein}
\abstracts{
Limits at the $10^{-4}$ level are reported for rare and forbidden decays of the
$D^0$ charm meson to a pair of leptons and either a vector meson
or two pseudoscalar mesons.  
Of these searches, 18 are investigations of decays without previous
published results; several others have significantly improved
sensitivity over previous results.
}{}{}

\textlineskip			
\vspace*{12pt}			

\noindent
We have made {\it blind} \, searches for 27 dilepton $D^0$ decays 
with muons or
electrons.\cite{rare4_791} 
The modes are resonant
$D^0\rightarrow V\ell^+\ell^-$ decays, where $V$ is a
$\rho ^0$, $\Kstar$, or $\phi$, and non-resonant 
$D^0\rightarrow hh\ell \ell $ decays, where $h$ is a $\pi$ or $K$.
Charge-conjugates are implied.
The modes are 
lepton flavor-violating, or
lepton number-violating, or
flavor-changing neutral current decays.
Box diagrams can mimic FCNC decays,
but only at the
$10^{-10}$ to $10^{-9}$ level.\cite{SCHWARTZ93,Fajfer} 
Long range effects (e.g.\ $D^{0}\rightarrow \Kstar \rho ^{0}, \
\rho^{0}\rightarrow  e^{+}e^{-}$) can occur at the $10^{-6}$ 
level.\cite{Fajfer,LD}
Many have studied rare decays of -1/3q strange
quarks.
Charge 2/3 charm quarks are interesting because they  might couple 
differently.\cite{Castro} 


Data is from the Fermilab E791
spectrometer.\cite{e791spect} 
The spectrometer has been upgraded for a series of experiments
including E516,\cite{e516} E691,\cite{e691} E769,\cite{e769} and E791.
In addition to searching for rare decays,\cite{rare4_791,FCNCnew} 
E791 has set limits
on $D^0 \, \overline{D}^{\,0}$ mixing\cite{mix} and {\it CP} violation.\cite{CP}
E791 has observed doubly cabibbo suppressed decays\cite{cabibbo} and has 
tagged quarks as being $c$ or $\overline{c}$
using $D$-$\pi$ production correlations.\cite{Dpi}
E791 recorded $2 \times 10^{10}$ events,\cite{da791} which were
produced by a 500 GeV/$c$~ $\pi ^{-}$ beam
hitting a fixed target consisting of five thin, well-separated
disks. Tracks and vertices use hits in 23 silicon
microstrip\cite{e691} and 45 wire chamber planes. 
Kaon
ID employs two \cerenkov{} 
counters.\cite{Bartlett}
Electron ID uses transverse shower shape plus matching wire chamber
tracks to shower positions and energies in an electromagnetic 
calorimeter.\cite{SLIC} The probability to mis-ID a pion as an electron
was $\sim
0.8\%$. Muon ID employs two planes of
scintillation counters.\cite{muon} 
Data from $D^+\rightarrow \overline{K}^{*0} \mu^{+}\nu
_{\!_{\mu}}$ decays\cite{Chong} were used to set cuts.
The probability to
mis-ID a pion as a muon decreased from about 6$\%$ at 
8~GeV/$c$ to $1.3\%$ above 20 GeV/$c$.

After reconstruction,\cite{farm791} events with well-separated
production and decay vertices were chosen to
separate charm candidates from background. 
All events having masses
near the $D^{0}$ mass were
{\it masked} so that any potential signal
would not cause bias. All cuts
were then chosen by studying signal events from Monte Carlo (MC)
and background events, outside signal
windows, from real data. 
We normalize to topologically similar
hadronic 3-body (resonant) or 4-body (non-resonant) decays. 
\begin{table}[t!]
\vspace*{-5pt}
\tabcolsep=1.5mm
\begin{center}
\begin{tabular}{lllllll}
\hline
\vspace*{-10pt} &  &  & & & & \\
$D^0$ Decay & $D^0$ Norm. & Events & \hspace*{5mm}&
$D^0$ Decay & $D^0$ Norm. & Events \\
\hline
\vspace*{-10pt} &  &  & & & & \\
$\rho^{0} \ell ^{\pm }\ell ^{\mp }$& $\pi ^+\pi ^-\pi ^+\pi ^-$& 2049$\pm$53 &&
$\pi \pi \ell\ell $& $\pi ^+\pi ^-\pi ^+\pi ^-$& 2049$\pm$53\\
$\Kstar \ell ^{\pm }\ell ^{\mp }$& $\Kstar \pi ^+\pi ^-$& 5451$\pm$72 &&
$K\pi \ell\ell $& $K^-\pi ^+\pi ^-\pi ^+$&11550$\pm$113 \\
$\phi \ell ^{\pm }\ell ^{\mp }$& $\phi \pi ^+\pi ^-$& 113$\pm$19 &&
$KK\ell\ell $& $K^+K^-\pi ^+\pi ^-$& 406$\pm$41\\
\hline
\end{tabular}
\end{center}
\vspace*{-5pt}
\end{table}
The upper limit for each branching fraction $B_{X}$ is calculated using
\begin{equation}
\vspace*{-5pt}
B_{X}=\frac{N_{X}}{N_{\mathrm{Norm}}}
\frac{\varepsilon _{\mathrm{Norm}}}{\varepsilon _{X}}
\times B_{\mathrm{Norm}}
\hspace*{15mm}
\frac{\varepsilon _{\mathrm{Norm}}}{\varepsilon _{X}}=
\frac{N_{\mathrm{Norm}}^{\mathrm{MC}}}{N_{X}^{\mathrm{MC}}}
\end{equation}
where $N_{X}$ is the 90$\%$ confidence level (CL) upper limit on the
number of decays for the rare or forbidden decay mode $X$, and
$\varepsilon_{X}$ is that mode's detection efficiency.
$\varepsilon_{\mathrm{Norm}}$ is the normalization mode detection
efficiency, $B_{\mathrm{Norm}}$ is the normalization mode branching
fraction, and
$N_{\mathrm{Norm}}^{\mathrm{MC}}$ and $N_{X}^{\mathrm{MC}}$ are
the numbers of Monte Carlo events that are reconstructed and pass
the final cuts, for the normalization and decay modes,
respectively. 
The efficiencies for the normalization modes varied from
$0.2\%$ to $1\%$, and the efficiencies for the
search modes varied from $0.05\%$ to $0.34\%$. 


The 90$\%$ CL upper limits $N_{X}$ are calculated using the method of
Feldman and Cousins\cite{Cousins} to account for background, and then
corrected for systematic errors by the method of Cousins and 
Highland.\cite{COUSINSHI}
Results are shown in
Fig.\ 1. 

\begin{figure}[ht!]
\begin{center}
\vspace*{-5pt}
\centerline{\epsfxsize 4.3 truein \epsfbox{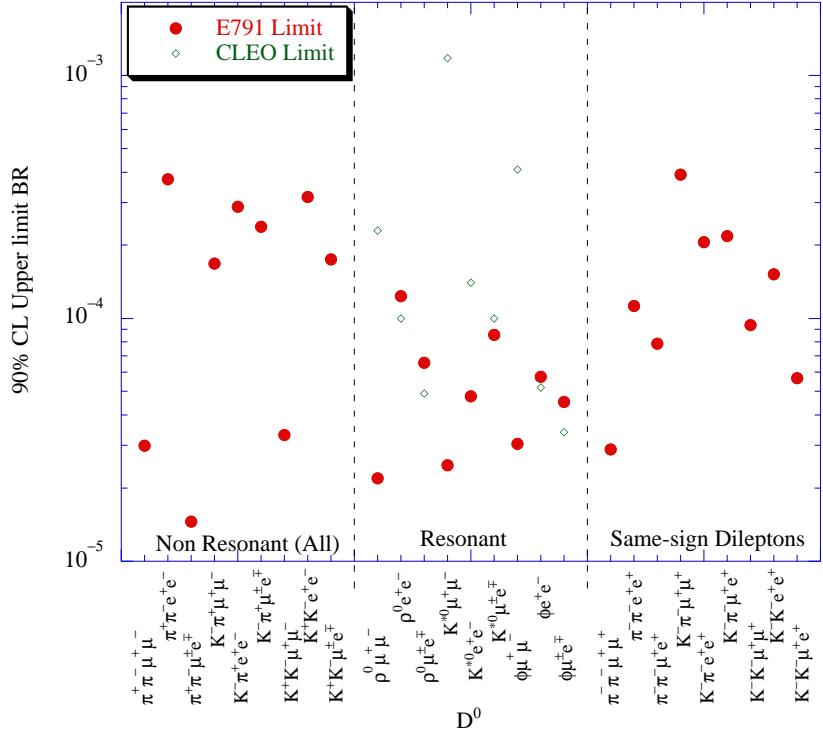}}
\vspace*{-10pt}
\caption[]{\small
\hbox{90\% CL branching fraction limits from E791\cite{rare4_791} ($\bullet$) \ 
and 
CLEO\cite{CLEO}} ($\diamond$).}
\end{center}
\end{figure}

\noindent
In summary, we use {\it blind} analyses to set 90\% CL upper limits 
ranging from $1.5 \times 10^{-5}$ ($\pi^+ \pi^- \mu^{\pm} e^{\mp}$) \,
to $3.9 \times 10^{-4}$ ($K^- \pi^- \mu^+ \mu^+$) \,
for
27 FCNC and lepton-number/ family violating decays of the
$D^0$.
No evidence for any of these 3 and 
4-body decays is
found.\cite{rare4_791}
Five limits are improvements over previous results;\cite{CLEO} 18 are
new.

\nonumsection{References}

\end{document}